\documentclass{ieeeaccess}
\usepackage{cite}
\usepackage{amsmath,amssymb,amsfonts,amsthm}
\usepackage{bbm}
\usepackage{mathtools}
\usepackage{physics}
\usepackage{algorithmic}
\usepackage{enumitem}
\usepackage{empheq}
\usepackage{multirow}
\usepackage{hyperref}
\usepackage{subcaption}
\usepackage{graphicx}
\usepackage{textcomp}
\usepackage{nicefrac}
\usepackage{accents}

\newtheorem{proposition}{Proposition}
\newtheorem{lemma}{Lemma}
\newtheorem{theorem}{Theorem}

\newcommand{\lbar}[1]{\underaccent{\bar}{#1}}

\def\BibTeX{{\rm B\kern-.05em{\sc i\kern-.025em b}\kern-.08em
    T\kern-.1667em\lower.7ex\hbox{E}\kern-.125emX}}
\begin{document}
\history{Date of publication xxxx 00, 0000, date of current version xxxx 00, 0000.}
\doi{XX.XXXX/XXX.202X.DOI}

\title{Convexification of the Quantum Network Utility Maximisation Problem}
\author{\uppercase{Sounak Kar}\authorrefmark{1} %\IEEEmembership{Fellow, IEEE},
\uppercase{and Stephanie Wehner\authorrefmark{1}}}
\address[1]{\{QuTech; Kavli Institute of Nanoscience; Quantum Computer Science, Department of Software Technologies, EEMCS\}, TU Delft, the Netherlands}
\address{emails: s.kar-1@tudelft.nl, s.d.c.wehner@tudelft.nl}
\tfootnote{This work was supported in part by NWO VICI grant VI.C.222.029.}

\markboth
{Kar \headeretal: Convexification of the Quantum Network Utility Maximisation Problem}
{Kar \headeretal: Convexification of the Quantum Network Utility Maximisation Problem}

%\corresp{Corresponding author: First A. Author (email: author@ boulder.nist.gov).}

\begin{abstract}
Network Utility Maximisation (NUM) addresses the problem of allocating resources fairly within a network and explores the ways to achieve optimal allocation in real-world networks. 
Although extensively studied in classical networks, NUM is an emerging area of research in the context of quantum networks.
In this work, we consider the quantum network utility maximisation (QNUM) problem in a static setting, where a user's utility takes into account the assigned quantum quality (fidelity) via a generic entanglement measure as well as the corresponding rate of entanglement generation.
Under certain assumptions, we demonstrate that the QNUM problem can be formulated as an optimisation problem with the rate allocation vector as the only decision variable.
Using a change of variable technique known in the field of geometric programming, we then establish sufficient conditions under which this formulation can be reduced to a convex problem— a class of optimisation problems that can be solved efficiently and with certainty even in high dimensions.
We further show that this technique preserves convexity, enabling us to formulate convex QNUM problems in networks where some routes have certain entanglement measures that do not readily admit convex formulation, while others do.
This allows us to compute the optimal
resource allocation in networks where heterogeneous applications run over different routes.
\looseness = -1
\end{abstract}


\titlepgskip=-15pt

\maketitle

\section{Introduction}\label{sec:intro}
\everypar{\looseness=-1}
\PARstart{Q}{uantum} networks are envisaged to facilitate a variety of applications, including quantum key distribution (QKD)~\cite{bennett1984quantum,ekert1992quantum}, enhanced sensing~\cite{giovannetti2004quantum,jozsa2000quantum} and blind quantum computation~\cite{broadbent2009universal,fitzsimons2017unconditionally}.
Unlike classical networks, where an application's quality of service (QoS) typically depends on the available transmission rate, the QoS of a quantum network application relies on the quality of entanglement and the rate at which it is distributed between the sender and the receiver.
Further, the QoS metric varies according to the underlying application, and the dependence of the QoS metric on the quality of entanglement can sometimes be captured via a suitable entanglement measure~\cite{vardoyan2023quantum}.
%Various entanglement measures, such as entanglement of formation, distillable entanglement and negativity, have been proposed to capture different aspects of entanglement quality, reflecting the diverse requirements across different quantum technologies~\cite{plenio2005introduction}. 

To support diverse applications and multiple users, a network must plan and distribute resources accordingly.
Two central concepts of resource distribution in networks are efficiency and fairness~\cite{leboudec2005rate}, where the latter has been the primary focus of Network Utility Maximisation (NUM)~\cite{kelly1997charging,kelly1998rate} in classical networks.
Since NUM is essentially a resource distribution problem, it borrows the mathematical framework of fairness from welfare economics~\cite{johansson1991introduction}, which explores the notion of \textit{equitable} resource allocation among contenders.
The core idea is to formulate a social welfare metric that takes into account the well-being of individuals.
Mathematically, this entails encoding individual well-being via suitable utility functions and aggregating them into a single social metric.
Subsequently, the social metric is maximised over possible resource allocations to find the optimal resource distribution.
Interestingly, there are social welfare metrics for which the optimal allocation is not necessarily Pareto-efficient, i.e., starting from the optimal allocation, it is possible to increase an individual's utility without affecting others~\cite{rawls1971theory}. 

In classical NUM, the transmission rate is usually the resource allocated across routes, defined as paths on the network graph.
To do so, the utility function of a route is formulated according to the QoS metric of the underlying application, such as delay, jitter or throughput.
Individual utilities are then aggregated into the (social) network utility function, which is optimised with respect to rate allocation.
Often, the individual utilities are concave functions of the rate allocation, enabling a central entity with the knowledge of individual utilities to compute the optimal rate allocation by solving a convex optimisation problem.
Since the global optimum can already be found efficiently and with certainty for convex problems, classical NUM literature explores other aspects such as decentralised implementations and their stability, i.e., the convergence of such implementations to the optimal allocation vector~\cite{kelly1997charging,kelly1998rate,palomar2006tutorial}.

In contrast to classical NUM, the quantum network utility maximisation (QNUM) problem~\cite{vardoyan2023quantum} aims to maximise the aggregate utility of the network over achievable entanglement quality and generation rate.
This is because the utility of a route in this case involves both the generation rate and the quality, where the latter is encoded via an entanglement measure~\cite{plenio2005introduction} (including the secret key rate) that depends on the underlying application.
Moreover, the rate and quality of entanglement generation are related- the relation being governed by the physical attributes of the quantum communication links. 
It was shown in~\cite{vardoyan2023quantum} that the aggregate utility function is not necessarily convex in the rate and quality of entanglement generation, meaning there is no theoretical guarantee for finding a globally optimal allocation.

In light of the above, this paper aims to find conditions under which the QNUM problem leads to a convex formulation.
Our setup differs from~\cite{gauthier2023architecture} in that their focus is on sharing the resources of an entanglement generation switch at the centre of a star network, while we are concerned with sharing link-level resources in a general topology.
A related but different problem on network planning was considered in~\cite{pouryousef2023quantum}, where network utility is maximised with respect to repeater locations in a network.
In contrast, our work assumes that the routes and the topology are fixed and only deals with distributing link-level resources to the routes.
We also slightly generalise the setup from~\cite{vardoyan2023quantum} by allowing the routes to have different entanglement measures to describe respective utilities.

Our first observation is that the QNUM problem can be reformulated as an optimisation problem with the entanglement generation rate as the sole decision variable in certain networks.
Borrowing a change of variable technique from geometric programming~\cite{boyd2007tutorial}, we then provide sufficient conditions for the QNUM problem to allow for convex reformulation.
Using the fact that the reformulation preserves convexity, we show that the QNUM problem can be transformed into a convex problem in the presence of certain route entanglement measures, some of which do not immediately lead to convex formulations while the rest do.
Our result has the implication that the optimal resource allocation for such QNUM problems can be computed efficiently and with certainty, even in large networks.

The remainder of the paper is structured as follows: in Sect.~\ref{sec:formulation}, we formally describe the setup and state our main results.
In Sect.~\ref{sec:examples}, we show that certain entanglement measures satisfy the conditions laid out in Sect.~\ref{sec:formulation}, while
the results are applied to an example network %to find the optimal allocation 
in Sect.~\ref{sec:eval}.
Finally, we present the proofs of the results in Sect.~\ref{sec:proofs}. \vspace{-0pt}

\section{Assumptions and Main Results}\label{sec:formulation}
\everypar{\looseness=-1}
\subsection{Setup and Assumptions}\label{sec:assumptions}
\begin{figure}[t]
\centering
\includegraphics[width=1.01\columnwidth]{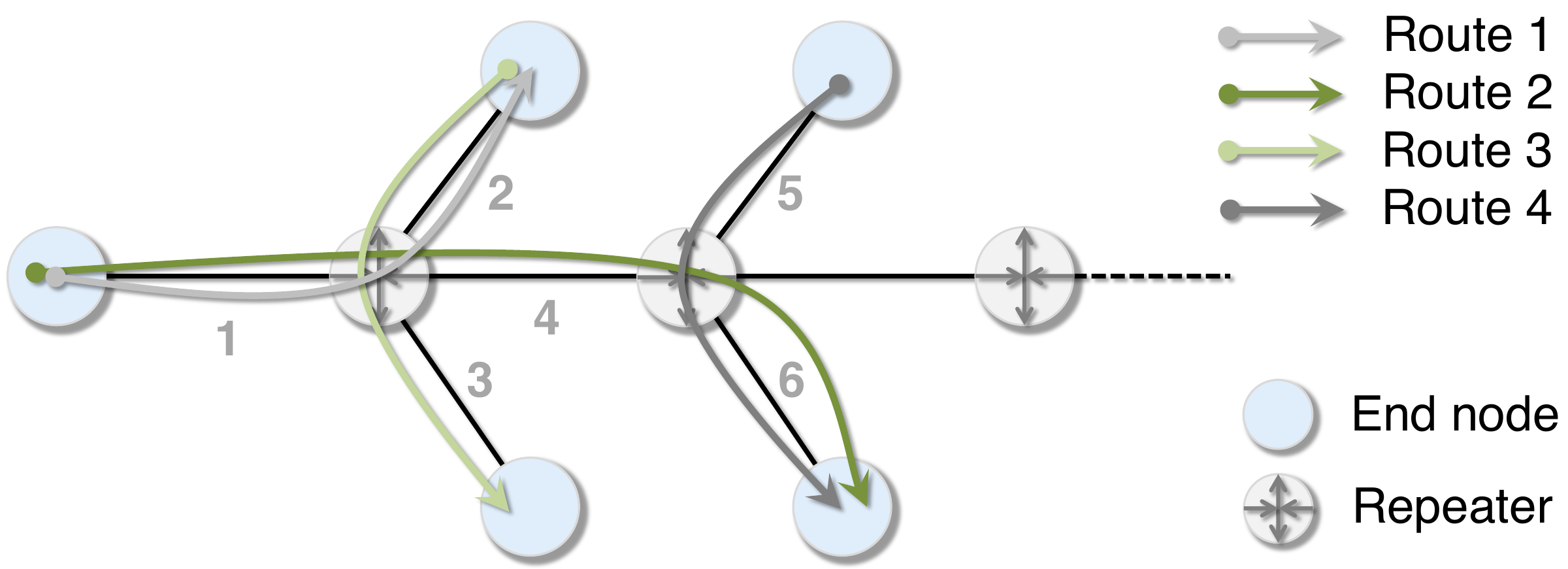}
\caption{\label{fig:net}%
An entanglement distribution network with numbered links: link $j$'s Werner parameter is $w_j,~j \in [6]$. The end-to-end (e2e) Werner parameters of the routes are products of corresponding link-level $w_j$'s. E.g., route $1$ and $2$ have e2e Werner parameters $w_1 w_2$ and $w_1 w_4 w_6$, respectively. The positive rate allocations $(x_1,x_2,x_3,x_4)$ must satisfy $6$ constraints, one for each link~\eqref{eq:QNUMcanonical}. For example, $x_4 \!\le\! d_5(1-w_5)$ (link $5$) and $x_2+x_4 \!\le\! d_6(1-w_6)$ (link $6$). The utility of a route is the product of the allocated rate and a measure of e2e entanglement, e.g., route $2$ has utility $x_2 f_2(w_1 w_4 w_6)$.}%
\vspace{-10pt}
\end{figure}

We consider an entanglement distribution network with a setup identical to~\cite{vardoyan2023quantum}.
The end nodes on this network represent users and they are connected via the repeater nodes.
Two nodes are adjacent iff there is a direct quantum communication link between them.
Further, a route is a path between two users, i.e., a sequence of adjacent nodes linking the corresponding end nodes.
For an example, see Fig.~\ref{fig:net}.
To describe and analyse the QNUM problem, we make the following assumptions: 
\begin{enumerate}[label={A\arabic*},nolistsep,leftmargin=*]
\item \label{A:entanglementSwap} \textbf{Entanglement swapping:} Entanglements between the end nodes are generated in two steps: first at the link level, i.e., between two adjacent nodes.
Next, entanglement swaps are performed at the repeater nodes along the route, producing end-to-end entanglement~\cite{pan1998experimental}.
\item \label{A:staticNet} \textbf{Static network:} Our goal is to distribute resources, i.e., entanglement generation rate and quality, between routes connecting end users who are running different applications.
Similar to NUM in its basic form, the routes and the applications are fixed.
This implies that the utility of a route changes only when its share of resources is modified.
%Formally, a route is a sequence of adjacent links of the network.
To solve the QNUM problem for a given network, we focus \textit{only} on the sub-network $\mathcal{G}$ consisting of relevant routes and the corresponding links.
We assume that $\mathcal{G}$ has $r$ routes and $l$ links.
%Note that this is akin to considering the NUM problem in its most basic form.
%
\item \label{A:singlePhoton} \textbf{Link level entanglement generation scheme:} We further assume that link-level entanglements are generated using the single-photon scheme~\cite{humphreys2018deterministic}, where the generated state has the following form:
\begin{align}\label{eq:singlePhoton}
    \rho = (1 - \alpha) \lvert \Psi^+ \rangle \langle \Psi^+ \rvert +  \alpha \lvert \uparrow \uparrow \rangle \langle \uparrow \uparrow \rvert~.    
\end{align}
Here $\alpha$ is the bright-state
population and $\ket{\Psi^+}$ is a Bell-state orthogonal to the bright state $\ket{\uparrow \uparrow}$.
Further, each generation attempt succeeds with probability
$$p_{\text{elem}} = 2 \kappa \eta \alpha~,\vspace{-0pt}$$
where $\eta$ denotes the transmissivity between one end of the link and its midpoint, and $\kappa \in (0,1)$ is a constant reflecting the inefficiencies other than photon loss in the fibre.
For a link of length $L$ km, its transmissivity is given by $\eta = 10^{-0.02 L}$.
\item \label{A:wernerState} \textbf{Entangled state description:} It is known that for networks, where link-level entanglements are generated as Werner states, the end-to-end Werner parameter of a route is given by the product of the Werner parameters of constituent links~\cite{munro2015inside}.
Also, any state can be transformed into a Werner state of the same fidelity by introducing depolarising noise~\cite{dur2007entanglement}.
Motivated by these two facts, we adopt a common worst-case model of Werner states to describe the elementary link states instead of~\eqref{eq:singlePhoton}:
\begin{align}\label{eq:werner}\vspace{-0pt}
    \rho_w = w \lvert \Psi^+ \rangle \langle \Psi^+ \rvert +  (1-w)~ \nicefrac{\mathbb{I}_4}{4}~. \vspace{-0pt}
\end{align}
Equating fidelities from~\eqref{eq:singlePhoton} and ~\eqref{eq:werner},
$$\vspace{0pt}1-\alpha = \frac{1+3w}{4} \implies \alpha = \frac{3(1-w)}{4}~.$$
\item \label{A:attemptFreq} \textbf{Frequency of entanglement generation attempt:} We assume that a link attempts entanglement generation every $T$ units of time, resulting in an entanglement generation rate of 
\begin{align}\label{eq:genRate}\vspace{-8pt}
    \frac{p_{\text{elem}}}{T} = \underbrace{\frac{3 \kappa \eta}{2T}}_{=:d}(1 - w)~.\vspace{-3pt}
\end{align}
Eq.~\eqref{eq:genRate} describes the relation between the rate of entanglement generation and the fidelity of the produced entanglement.
For link $j \in [l]$, if we fix the fidelity of the produced entanglement by fixing the Werner parameter as $w_j$, the maximum rate at which link $j$ will be able to produce entanglement is given by $\mu_j := d_j(1-w_j)$, where $d_j := 3 \kappa_j \eta_j/ 2T_j$.
Note that $\mu_j$ can be thought of as the capacity of link $j$ when it produces Werner states with parameter $w_j$.
\item \label{A:fixedFidelity} \textbf{Arbitrary but fixed quality of entanglement:} The Werner parameter of a link-level entanglement $w_j, j\in [l]$ can be chosen arbitrarily for optimising the network utility but remains fixed once chosen.
That is, the contributions of the $j$-th link towards the end-to-end Werner parameters are the same across the routes passing through that link.
%This requires setting the bright state parameter $\alpha$ suitably in~\eqref{eq:singlePhoton} for link-level entanglement generation.
Observe from~\eqref{eq:genRate} that increasing the value of the Werner parameter reduces the entanglement generation rate. 
\item \label{A:utility} \looseness = -1 \textbf{Utility of a route:} We denote the allocated rate of route $i$ by $x_i$ and the end-to-end Werner parameter by $u_i$.
To simplify the formulation, we also introduce the (binary) link-route incidence matrix $A$, where $a_{ji}\!:=\!((A))_{ji}\!=\!1$ iff the $i$-th route passes through the $j$-th link.
Note that we must have (i) $\sum_{i \in [r]}a_{ji} x_i \le \mu_j$, i.e., the total rate allocated to the incident routes cannot exceed a link's maximum entanglement generation rate and (ii) $u_i = \prod_{j \in [l]}w_j^{a_{ji}}$, i.e., the end-to-end Werner parameter is the product of link-level Werner parameters~\cite{munro2015inside}.
To reflect the suitability of a Werner state for executing the underlying application for the $i$-th route, we use an entanglement measure (including Secret Key Rate) $f_i$ where 
\begin{align*}\vspace{-5pt}
    f_i:&[0,1] \to [0,b_i] \\
    & u_i \mapsto f_i(u_i)\vspace{-3pt}
\end{align*}
We assume that $f_i$ is non-decreasing and twice differentiable on $\{z\!:\!f_i(z)\!>\!0\}\!\setminus\!\{1\}$, and $f_i(0) = 0$ for $i \in [r]$.
The utility of a route is assumed to have the form $x_i f_i(u_i)$.
Finally, the network utility is formulated as the product of the route utilities. 
The product form of individual and network utilities ensures that a specified level of network utility is achieved only when each route receives adequate rate and fidelity allocations.
Note that it is possible to have different forms for route and network utility functions than ours.
\end{enumerate}
Equipped with the assumptions, we are ready to describe the QNUM problem introduced in~\cite{vardoyan2023quantum}.
A complete list of parameters describing the network and the auxiliary variables are given in Tab.~\ref{tab:defs}.

\vspace{-0pt}
\begin{center}
\begin{tabular}{ c l }
  \hline
  \hline
  $[n]$ & $\{1,2,\dots,n\}$ for $n \in \mathbb{N}$ \\
  $l$ & the number of links in the network \\
  $r$ & the number of routes in the network \\
  $d_j$ & a positive constant characterising the physical \\
  & attributes of the $j$-th link,
  see~\eqref{eq:genRate} \\
  $w_j$ & the Werner parameter of the generated pairs in the\\
  & $j$-th link \\
  $\mu_j$ & the corresponding entanglement generation capacity \\
  &  of the $j$-th link, $\mu_j:= d_j(1-w_j)$ \\
  $x_i$ & the rate allocated to the $i$-th route\\
  $y_i$ & $\ln(x_i)$ \\
  $a_{ji}$ & the binary variable taking value $1$ iff the $i$-th route \\
  & passes through the $j$-th link\\
  $A$ & the link-route incidence matrix, i.e., $((A))_{ji} \!=\! a_{ji}$ \\
  $A_j$ & the $j$-th row of $A$, i.e.,  $(a_{j1},a_{j2},\dots,a_{jr})$\\
  $u_i$ & the end-to-end Werner parameter of the $i$-th route,\\
  & i.e., $u_i = \prod\limits_{j=1}^l w_j^{a_{ji}}$ \vspace{-0pt}\\ 
  $u_i(\vec{y})$ & $\prod\limits_{j=1}^l \big(1-\langle A_j,e^{\vec{y}} \rangle/d_j\big)^{a_{ji}}$, $\Vec{y} \!\in\! \mathbb{R}^r$\\
  $f_i$ & the entanglement measure for the $i$-th route, \\
  & $f_i\!:\![0,1] \!\to\! [0,b_i]$, non-decreasing and twice  \\
  & differentiable on $\{z\!:\!f_i(z)\!>\!0\}\!\setminus\!\{1\}$, and $f_i(0) \!=\! 0$ \\
  $c^{(i)}$ & $\sup \{z: f_i(z)=0\}$\\
  $T$ &$\{\Vec{y} \!\in\! \mathbb{R}^r\!\!:\langle A_j,e^{\vec{y}} \rangle \!<\!d_j ~ \forall j\}$\\
  $S_i$ & $\{\Vec{y} \!\in\! T\!:u_i(\Vec{y})\!>\!c^{(i)}\!\}$\\
  $F_i$ & $\ln f_i$,~ $F_i:(c^{(i)},1]\to \mathbb{R}$\\
  $c_1^{(i)}$ & unique inflection point of $F_i$, $F_i$ is concave in \\
  & $(c^{(i)},c_1^{(i)}]$ and convex in $(c_1^{(i)},1)$\\
  \hline
\end{tabular}
\captionof{table}{List of notations: the link index $j \!\in\! [l]$ and the route index $i \!\in\! [r]$.}
\vspace{-0pt}
\label{tab:defs}
\end{center}

\vspace{-3pt}
\subsection{The QNUM Problem}\label{sec:qnumProblem}
We denote the rate allocation vector for the routes by ${\vec{x} \!=\! (x_1,x_2,\dots,x_r)}$ and the Werner parameter vector for the links by ${\vec{w} \!=\! (w_1,w_2,\dots,w_l)}$.
The QNUM problem in its canonical form can then be written as
\begin{align}\label{eq:QNUMcanonical}\vspace{-0pt}
  \max\limits_{\vec{x},\vec{w}} \quad & \prod\limits_{i=1}^r x_i f_i\bigg(\prod\limits_{j=1}^l w_j^{a_{ji}}\bigg) \nonumber\\
  \text{s.t.} \quad & \vec{0} \prec \vec{x}~, \\
  & \vec{0} \prec \vec{w} \preceq \vec{1}~,\thickspace \text{(Fidelity bounds)} \nonumber\\
  &\langle A_j,\vec{x} \rangle \!\le\! \mu_j \!=\! d_j(1\!-\!w_j) \thinspace~ \forall j \!\in\! [l].\thickspace \text{(Rate constraints)}\nonumber %\vspace{-3pt}
\end{align}

Here, $\preceq$ (resp. $\prec$) denotes element-wise (resp. strict) inequality and $\langle A_j,\vec{x} \rangle$ denotes the dot product of $A_j$ and $\vec{x}$.
Further, the inequalities in $\vec{0}\! \prec\! \vec{x}$ and $\vec{0}\! \prec\! \vec{w}$ are strict as the objective function is non-negative and equals zero if any element of $\vec{x}$ or $\vec{w}$ is zero.
%The product form of the network utility function ensures that each route has reasonable allocation.
Note that for any feasible $(\vec{x},w_j)$, if the last inequality in~\eqref{eq:QNUMcanonical} is strict, i.e., $\langle A_j,\vec{x} \rangle \!<\!d_j(1\!-\!w_j)$, it is possible to increase $w_j$ further to make it an equality.
This is because, with all other variables held constant, an increase in \(w_j\) results in a higher or equal value of the objective function, as \(f_i\)'s are non-decreasing by assumption.
Thus, if there exists a solution to~\eqref{eq:QNUMcanonical}, there will be another solution satisfying
\begin{align}\label{eq:wFromX}\vspace{-2pt}
  \langle A_j,\vec{x} \rangle \!=\! d_j(1\!-\!w_j) \thinspace\implies\thinspace
  w_j \!=\! 1\!-\!\frac{\langle A_j,\vec{x} \rangle}{d_j} \thickspace \forall~j\! \in\! [l],\vspace{-2pt}
\end{align}
\looseness = -1 for which the objective function attains a higher or equal value.
Therefore, it is sufficient to focus only on solutions satisfying~\eqref {eq:wFromX}, which allows us to eliminate $\vec{w}$ from the set of optimisation variables.
Also, instead of maximising~\eqref{eq:QNUMcanonical}, we minimise the negative logarithm of the objective function, which leads to the following formulation:
\begin{align}\label{eq:QNUMtransformedInt}
    \min\limits_{\vec{x}} & -\!\!\sum\limits_{i=1}^r \!\bigg(\!\ln x_i \!+\! \ln\bigg(f_i\bigg(\prod\limits_{j=1}^l \bigg( 1\!-\!\frac{\langle A_j,\vec{x} \rangle}{d_j}\bigg)^{a_{ji}}\bigg)\!\bigg) \nonumber\\
    \text{s.t.}&\quad  \vec{0} \prec \vec{x}~, \nonumber\\
    & \quad 0<\frac{\langle A_j,\vec{x} \rangle}{d_j}<1~, \thickspace j \in [l]~, \\
    &\quad c^{(i)} <\prod\limits_{j=1}^l \bigg( 1-\frac{\langle A_j,\vec{x} \rangle}{d_j}\bigg)^{a_{ji}},\thickspace i \in [r]~. \nonumber
\end{align}
Here $c^{(i)}\!:=\! \sup \{z: f_i(z)\!=\!0\}$.

We now comment on the constraints of formulation~\eqref{eq:QNUMtransformedInt}.
The constraint $0\! \prec\! \vec{x}$, together with~\eqref{eq:wFromX}, modifies the constraint $\vec{0}\! \prec\! \vec{w}\! \preceq\! \vec{1}$ in~\eqref{eq:QNUMcanonical} to $0\!<\!w_j\!<\!1$, $\forall j \in [l]$, as reflected in the second constraint.
The last constraint guarantees that the logarithms in the objective function have positive arguments. 

In~\cite[App. B]{vardoyan2023quantum}, it was shown that the objective function of the QNUM problem is not necessarily convex in rate and quality allocations.
To convert~\eqref{eq:QNUMtransformedInt} into a convex problem, we perform the following change of variable well-known in the field of geometric programming~\cite{boyd2007tutorial}:
\begin{align}\label{eq:cov}
    \vec{x}=e^{\vec{y}}:= (e^{y_1},e^{y_2},\dots,e^{y_r})~.
\end{align}
We will show that this leads to a convex formulation under certain conditions.
Substituting~\eqref{eq:cov} in~\eqref{eq:QNUMtransformedInt} yields
\begin{align}\label{eq:QNUMtransformed}
    \min\limits_{\vec{y} \in \mathbb{R}^r} & -\sum\limits_{i=1}^r \bigg(y_i + \ln\bigg(f_i\bigg(\prod\limits_{j=1}^l \bigg( 1\!-\!\frac{\langle A_j,e^{\vec{y}} \rangle}{d_j}\bigg)^{a_{ji}}\bigg)\bigg) \nonumber\\
    \text{s.t.}&\quad  0<\frac{\langle A_j,e^{\vec{y}} \rangle}{d_j}<1~, \thickspace j \in [l]~, \\
    &\quad c^{(i)} <\prod\limits_{j=1}^l \bigg( 1-\frac{\langle A_j,e^{\vec{y}} \rangle}{d_j}\bigg)^{a_{ji}},\thickspace i \in [r]~.\nonumber
\end{align}
Here, the implicit constraint $\vec{y} \in \mathbb{R}^r$ is imposed to ensure $-\vec{\infty}\!\prec\!\vec{y}$, i.e., $0\! \prec\! \vec{x}$ as required in~\eqref{eq:QNUMtransformedInt}.
For brevity, we conveniently reuse the notations for $w_j$ and $u_i$:
%
% \begin{align}\label{eq:u}
%     u_i(\vec{y}) := \prod\limits_{j=1}^l \bigg( 1-\frac{\langle A_j,e^{\vec{y}} \rangle}{d_j}\bigg)^{a_{ji}}~.
% \end{align}
% %
% We also conveniently reuse the notation for $w_j$
%
\begin{align}\label{eq:wFromY}
    w_j(\vec{y}) := 1\!-\!\frac{\langle A_j,e^{\vec{y}} \rangle}{d_j}~, \quad u_i(\vec{y}) := \prod\limits_{j=1}^l \big(w_j(\vec{y})\big)^{a_{ji}}~.
\end{align}

In order for~\eqref{eq:QNUMtransformed} to be a convex optimisation problem, we need its objective function and the feasible region to be convex.
To describe the feasible region, we define
\begin{align}\label{eq:feasibleReg}\vspace{-5pt}
\begin{aligned}
    T_j \!:= &\{\Vec{y} \!\in\! \mathbb{R}^r\!\!:\langle A_j,e^{\vec{y}} \rangle \!<\!d_j\}, \thickspace j \in [l], ~\thickspace 
    T \!:= \bigcap\limits_{j \in [l]}^{} T_j~, \\
    S_i \!:= &\{\Vec{y} \!\in\! T\!\!:u_i(\Vec{y})\!>\!c^{(i)}\!\}, \thickspace i \in [r], ~\thickspace
    S \!:= \bigcap\limits_{i \in [r]}^{} S_i~.
\end{aligned}\vspace{-5pt}
\end{align}
Since $\langle A_j,e^{\vec{y}} \rangle \!>\!0$ already holds for $\Vec{y} \!\in\! \mathbb{R}^r$, $S$ is the feasible region for problem~\eqref{eq:QNUMtransformed}.
In the next subsection, we show that $S$ is a convex set.
We also provide sufficient conditions for the objective function to be convex.
\vspace{-0pt}
\subsection{Results}\label{sec:results}
While the feasible region of the transformed problem~\eqref{eq:QNUMtransformed} is always convex, the objective function is not in general.
We take the following approach to derive the conditions for it to be convex: we consider the contributions from each route to the objective function (i.e., $-y_i\!-\!\ln f_i(u_i(\vec{y})),~i \!\in\! [r]$) and look for conditions for them to be convex individually.
Since a sum of convex functions is convex, the objective function is convex if all individual conditions are satisfied.
The individual conditions are provided as (Cond.~1) and (Cond.~2) in~\eqref{eq:c>.5} and~\eqref{eq:condFconcave}, respectively.
We also show that the change of variable~\eqref{eq:cov} preserves convexity, i.e., if the contribution of a route to the objective function in~\eqref{eq:QNUMtransformedInt} is convex, so is the corresponding contribution in the reformulation~\eqref{eq:QNUMtransformed}.
This allows us to apply technique~\eqref{eq:cov} to convexify the contribution from a route without affecting the behaviour of already convex contributions from other routes. 

\begin{theorem}\label{prop:convexDomain}
  The transformed QNUM problem~\eqref{eq:QNUMtransformed} is feasible, and the set of feasible vectors $S$ is a convex set.
\end{theorem}
\begin{proof}[Proof idea]
  The problem is feasible since it is possible to satisfy constraints in~\eqref{eq:QNUMtransformed} by allocating sufficiently small rates to each route, i.e., by taking $\vec{y} \preceq -M \vec{1}$ for sufficiently large $M\!>\! 0$.
  Convexity of $S$ follows from convexity of each $S_i$ and $T_j$.
  See Sect.~\ref{sec:proofs} for a complete proof. \vspace{-5pt}
\end{proof}

\looseness = -1 To establish the convexity of the objective function on $S$, we first note that $-y_i$ is a convex function of $\vec{y}$.
Since a sum of convex functions is convex, we only look for a sufficient condition for $-\ln(f_i( u_i(\vec{y})))$ to be convex, i.e., for $\ln(f_i( u_i(\vec{y})))$ to be concave on $S_i$, defined in~\eqref{eq:feasibleReg}.
The following proposition, which provides a sufficient condition for $u_i(\vec{y})$ to be concave on $S_i$, is a stepping stone towards that goal. 
\begin{proposition}
   For $i \in [r]$, let $H^{(i)}(\vec{y})$ denote the Hessian of $u_i(\vec{y})$.
   If 
   \begin{flalign}\label{eq:c>.5}
     \textup{(Cond.~ 1)}\quad c^{(i)}:= \sup \{z: f_i(z)=0\} \ge \nicefrac{1}{2}~,
   \end{flalign}
   $H^{(i)}(\vec{y})$ is negative semidefinite on $S_i$.
\label{prop:uiConcave}
\end{proposition}
\begin{proof}[Proof idea]
  We compute the Hessian and show that its eigenvalues are non-positive on $S_i$ if ${c^{(i)} \!\ge\! 1/2}$.
  See Sect.~\ref{sec:proofs} for details.
\end{proof}
Our main result is the following:
\begin{theorem}\label{thm:FuConcave}
  Let Cond. 1~\eqref{eq:c>.5} hold and $f_i$ be twice differentiable on $(c^{(i)},1)$. Assume that ${F_i(u) \!:=\! \ln f_i(u),~u \in (c^{(i)},1]}$ has a unique inflection point $c_1^{(i)}\!\ge\! c^{(i)}$ satisfying ${F_i''(u)\!\le\! 0, ~\forall u\!\in\! (c^{(i)},c_1^{(i)}]}$ and ${F_i''(u)\!>\!0, ~\forall u\!\in\! (c_1^{(i)},1)}$.
  Further, let
  \begin{align}\label{eq:condFconcave}
    \textup{(Cond.~ 2)}\quad v_i(u) \!:=\! \frac{u F_i^{\prime \prime}(u)}{u F_i^{\prime \prime}(u) \!+\! F_i^{\prime}(u)}\!+\!\frac{1}{u} \!\le\! 2, \thickspace \forall u \!\in\! (c_1^{(i)}\!,1)~.
  \end{align}
  Then, $F_i(u_i(\vec{y}))$ is concave on $S_i$.
\end{theorem}
\begin{proof}[Proof idea]
  Since $u_i$ is concave on $S_i$ and $F_i$ is concave and increasing on $(c^{(i)}\!,c_1^{(i)}]$, $F_i(u_i(\vec{y}))$ is also concave on ${\{\Vec{y} \!\in\! T\!\!:\!c^{(i)}\!\!<\!u_i(\Vec{y})\!\le\!c_1^{(i)}\}}$.
  On $\{\Vec{y} \!\in\! T\!\!:\!c_1^{(i)}\!\!<\!u_i(\Vec{y})\!\!<\!1\}$, we show that the eigenvalues of its Hessian are non-positive if Cond. 1~\eqref{eq:c>.5} and Cond. 2~\eqref{eq:condFconcave} are satisfied.
  See Sect.~\ref{sec:proofs} for details. 
\end{proof}

We now show that the change of variable~\eqref{eq:cov} preserves convexity.
To that end, we note that the contribution from the $i$-th route to the objective function in formulation~\eqref{eq:QNUMtransformedInt} is \vspace{-0pt}
\begin{align}\label{eq:G}
\begin{aligned}
    h_x(\vec{x}) &:= -\ln x_i \!-\!G(x)~, \quad \text{where}\\ 
    G(x) &:=F_i\Big(\prod\limits_{j=1}^l \Big( 1\!-\!\frac{\langle A_j,\vec{x} \rangle}{d_j}\Big)^{a_{ji}}\Big)~.    
\end{aligned}\vspace{-12pt}
\end{align}
The corresponding contribution in formulation~\eqref{eq:QNUMtransformed} is \vspace{-2pt}
$$h(\vec{y}) := -y_i-F_i(u_i(\vec{y}))~.\vspace{-2pt}$$
The following proposition formalises our argument.
\begin{proposition}\label{prop:FxToFy}
   If $h_x(\vec{x})$ is convex, so is $h(\vec{y})$, i.e., the change of variable in~\eqref{eq:cov} preserves convexity.
\end{proposition}
\begin{proof}[Proof]
  Since $-\!\ln x_i$ is a convex function of $\vec{x}$, we essentially show that if $G(\vec{x})$ is concave, so is $F_i(u_i(\vec{y}))$.
  Note that $F_i(u_i(\vec{y})) = G(e^{\vec{y}})$.
  Let us denote the Hessian of $G(\vec{y})$ and $F_i(u_i(\vec{y}))$ by $D^2G(\vec{y})$ and $D^2F(\vec{y})$, respectively.
  Then,
  \begin{align*}\vspace{-5pt}
      D^2F(\vec{y}) &= E(\vec{y})D^2G(e^{\vec{y}})E(\vec{y})+E(\vec{y})\nabla G(e^{\vec{y}})\quad \text{where}\\
      E(\vec{y}) &:= \text{diag}(e^{y_1},e^{y_2},\dots,e^{y_r})~,\\
      \nabla G(\vec{y}) &:= \text{diag}\Big(\frac{\partial G}{\partial y_1}(\vec{y}),\frac{\partial G}{\partial y_2}(\vec{y}),\dots,\frac{\partial G}{\partial y_r}(\vec{y})\Big)~.\vspace{-5pt}
  \end{align*}
  Since $F_i$ is non-decreasing and $0\!<\!1\!-\!\langle A_j,\vec{x} \rangle/{d_j}\!<\!1$ on our domain of interest, observe from~\eqref{eq:G} that each $\partial G/\partial y_k$ is non-positive for $k \in [r]$.
  Thus, $E(\vec{y})\nabla G(e^{\vec{y}})$ is negative semidefinite.
  Also, by assumption, $h_x(\vec{x})$ is convex, i.e., $G(\vec{x})$ is concave, which implies that $D^2G$ is negative semidefinite. Hence, $E(\vec{y})D^2G(e^{\vec{y}})E(\vec{y})$ is negative semidefinite as well.
  This proves that $D^2F$ is negative semidefinite, i.e., $F_i(u_i(\vec{y}))$ is concave, as required.
\end{proof}

\looseness = -1 We have thus established that if each utility function $f_i$ either satisfies Cond. 1 and Cond. 2 or its contribution~\eqref{eq:G} to the objective function in formulation~\eqref{eq:QNUMtransformedInt} is already convex, formulation~\eqref{eq:QNUMtransformed} is a convex optimisation problem.
We now test these criteria on certain entanglement measures.
\section{Example Entanglement Measures}\label{sec:examples}

\everypar{\looseness=-1}
We first show that the entanglement measures considered in~\cite{vardoyan2023quantum} that did not \emph{readily} admit convex formulations satisfy Cond. 1~\eqref{eq:c>.5} and Cond. 2~\eqref{eq:condFconcave}, and thus can be transformed into a convex problem via formulation~\eqref{eq:QNUMtransformed}.
We then provide an example where the entanglement measure does not satisfy Cond. 1~\eqref{eq:c>.5} but satisfies the hypothesis of Prop.~\ref{prop:FxToFy}, i.e., convexity of the contribution towards the objective function is preserved by the change of variable ~\eqref{eq:cov} for routes using this measure of entanglement.
We end with an example where the entanglement measure satisfies none of the aforementioned conditions but admits a convex formulation once we impose a cut-off on the end-to-end Werner parameters. 

\begin{figure*}[t]
\centering
\begin{subfigure}{0.31\textwidth}
    \centering
    \includegraphics[width=1\textwidth]{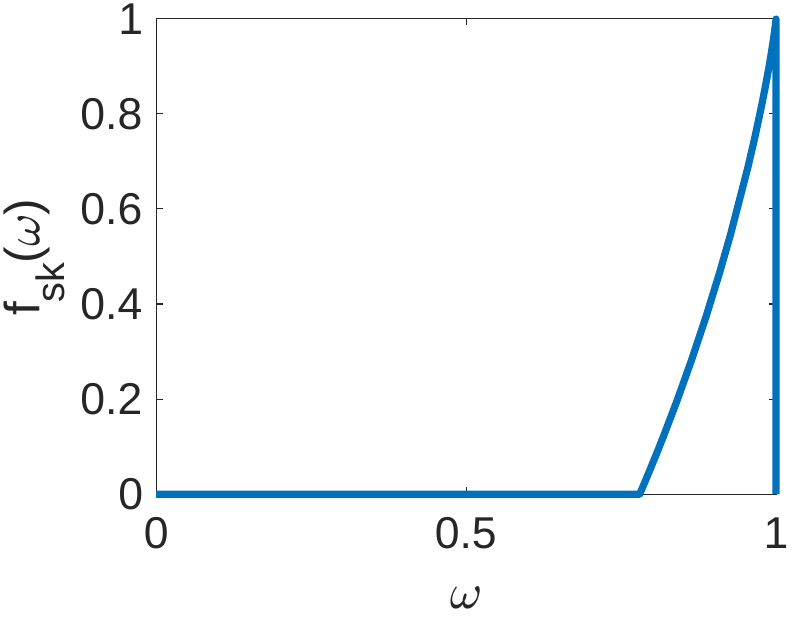}
    \caption{\label{fig:f_sk}%
    Entanglement measure $f_\text{sk}$}
\end{subfigure}
    \hfill
\begin{subfigure}{0.31\textwidth}
    \centering
    \includegraphics[width=1\textwidth]{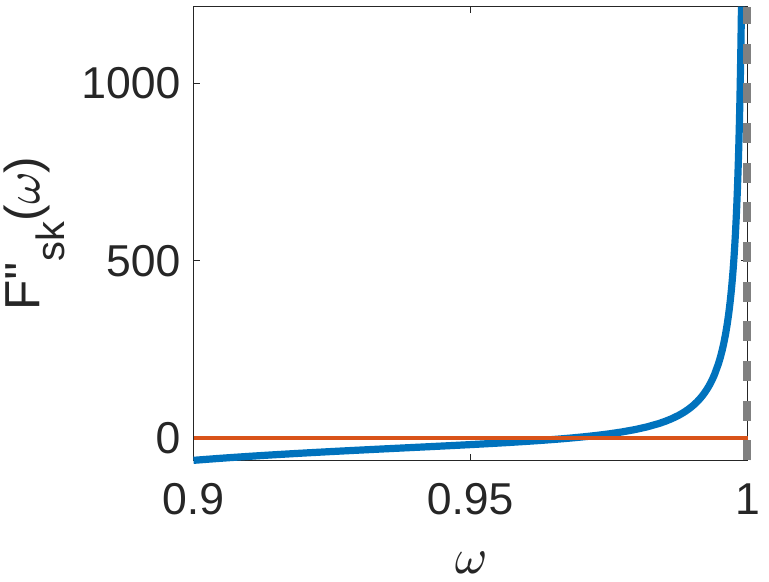}
    \caption{\label{fig:d2F_sk}%
    Second derivative of $\ln f_\text{sk}$}
\end{subfigure}
\hfill
\begin{subfigure}{0.31\textwidth}
    \centering
    \includegraphics[width=1\textwidth]{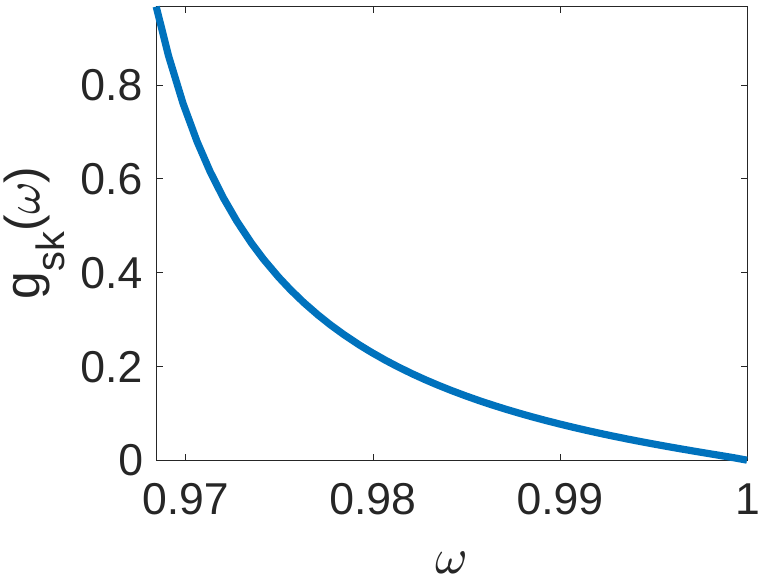}
    \caption{\label{fig:g_sk}%
    Condition~\eqref{eq:condFconcave}}
\end{subfigure}%
\caption{\label{fig:sk}%
(a) The secret key fraction satisfies $\sup \{\omega: f_\text{sk}(\omega)=0\} \approx 0.779944 \ge 1/2$. (b) The unique inflection point of its logarithm $F_\text{sk}$ is approximately $0.968418$. (c) For $\omega\!>\!0.968418$, we plot ${g_\text{sk}(u) \!=\! 2\!-\!u F_\text{sk}^{\prime \prime}(u)/(u F_\text{sk}^{\prime \prime}(u) \!+\! F_\text{sk}^{\prime}(u))\!-\!1/u}$ showing that Cond. 2~\eqref{eq:condFconcave} is satisfied.}%
\vspace{-8pt}
\end{figure*}

\begin{figure*}[t]
\centering
\begin{subfigure}{0.31\textwidth}
    \centering
    \includegraphics[width=1\textwidth]{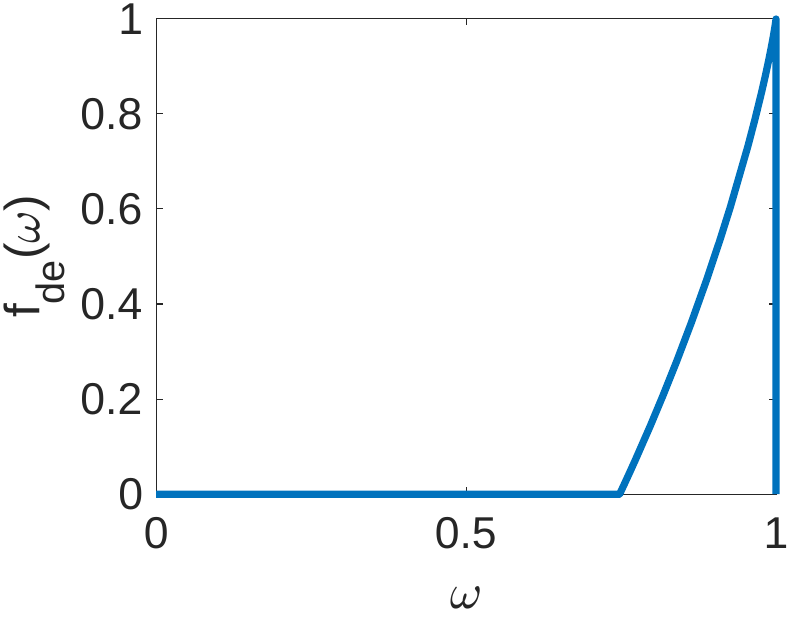}
    \caption{\label{fig:f_de}%
     Entanglement measure $f_\text{de}$}
\end{subfigure}
    \hfill
\begin{subfigure}{0.31\textwidth}
    \centering
    \includegraphics[width=1\textwidth]{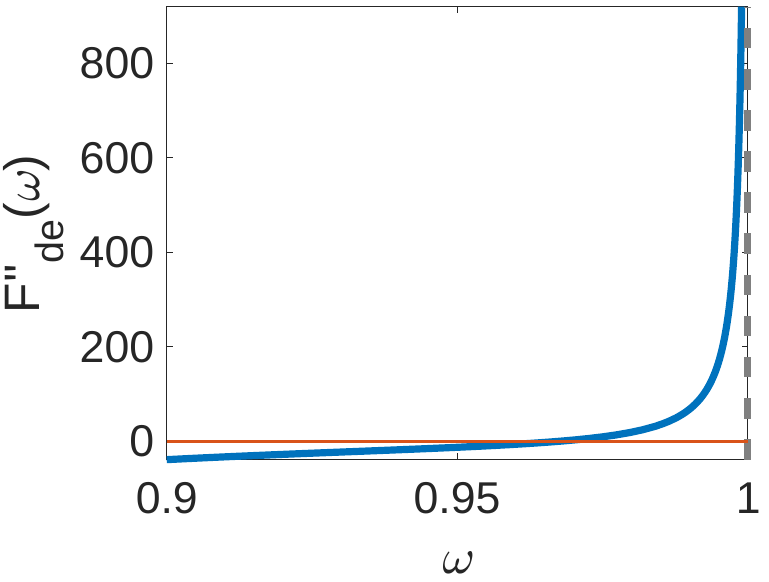}
    \caption{\label{fig:d2F_de}%
    Second derivative of $\ln f_\text{de}$}
\end{subfigure}
\hfill
\begin{subfigure}{0.31\textwidth}
    \centering
    \includegraphics[width=1\textwidth]{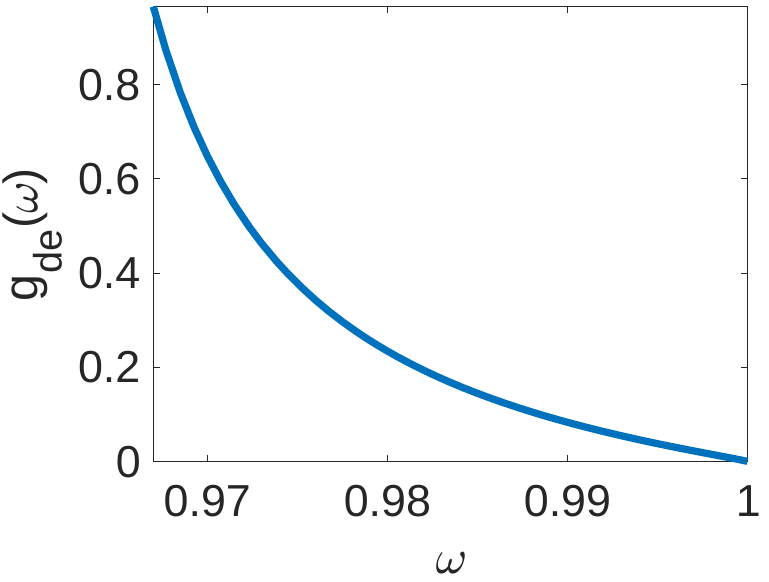}
    \caption{\label{fig:g_de}%
    Condition~\eqref{eq:condFconcave}}
\end{subfigure}%
\caption{\label{fig:de}%
(a) The lower bound to distillable entanglement~\eqref{eq:de} satisfies $\sup \{\omega: f_\text{de}(\omega)=0\} \approx 0.747613 \ge 1/2$. The unique inflection point of $F_\text{de}:=\ln(f_\text{de})$ is approximately $0.966984$ (b), beyond which Cond. 2~\eqref{eq:condFconcave} is shown to be satisfied by plotting ${g_\text{de}(u) \!=\! 2\!-\!u F_\text{de}^{\prime \prime}(u)/(u F_\text{de}^{\prime \prime}(u) \!+\! F_\text{de}^{\prime}(u))\!-\!1/u}$ in (c).}%
\vspace{-8pt}
\end{figure*}

\subsection{Secret Key Fraction}\label{sec:skf}

We first consider the secret key fraction~\cite{shor2000simple}, which has the following form for Werner states with Werner parameter $\omega$:
\begin{align}\label{eq:sk}
    f_\text{sk}(\omega) \!=\! \max \Big(0,1\!+\!(1\!+\!\omega) \log_2 \frac{1\!+\!\omega}{2}+ 
    (1\!-\!\omega) \log_2 \frac{1\!-\!\omega}{2} \Big).
\end{align}
Since, $f_\text{sk}(1/2)=0$, Cond.~1~\eqref{eq:c>.5} is satisfied.
In particular,
$$c^\text{sk}:=\sup \{\omega: f_\text{sk}(\omega)=0\} \approx 0.779944~.$$ 
For $\omega>c^\text{sk}$, we define $F_\text{sk} := \ln(f_\text{sk})$.
Also,
\begin{align*}
    f'_\text{sk}(\omega) &= \log_2 \Big( \frac{1\!+\!\omega}{1\!-\!\omega}\Big)~, \quad
    f''_\text{sk}(\omega) = \frac{2 \log_2 e}{1-\omega^2}~, \\
    F'_\text{sk}(\omega) &= \frac{f'_\text{sk}(\omega)}{f_\text{sk}(\omega)}~, \quad F''_\text{sk}(\omega) = \frac{f''_\text{sk}(\omega) f_\text{sk}(\omega)\!-\! (f'_\text{sk}(\omega))^2}{(f_\text{sk}(\omega))^2}~.
\end{align*} 
In Fig.~\ref{fig:d2F_sk}, we see that $F_\text{sk}$ has a unique inflection point $c_1^\text{sk} \!\approx\! 0.968418$. 
Further, Cond.~2~\eqref{eq:condFconcave} is seen to be true in Fig.~\ref{fig:g_sk}, where we plot ${g_\text{sk}(u) \!=\! 2\!-\!u F_\text{sk}^{\prime \prime}(u)/(u F_\text{sk}^{\prime \prime}(u) \!+\! F_\text{sk}^{\prime}(u))\!-\!1/u}$.
%Thus, by Thm.~\ref{prop:convexDomain}, the feasible region $S_i$ (see ~\eqref{eq:feasibleReg}) corresponding to a route with $f_i = f_\text{sk}$ is convex.
Thus, by Thm.~\ref{thm:FuConcave}, the contribution of a route to the objective function in formulation~\eqref{eq:QNUMtransformed} is convex if $f_\text{sk}$ is used as its measure of entanglement. \vspace{-7pt}

\subsection{Distillable Entanglement}\label{sec:de}
Following~\cite{vardoyan2023quantum}, we consider a lower-bound to distillable entanglement.
For a Werner state with Werner parameter $\omega$, the lower bound can be expressed as
\begin{align}\label{eq:de}\vspace{-5pt}
    f_\text{de}(\omega) &\!=\! \max \Big(0,1+\frac{1+3\omega}{4}\log_2\Big(\frac{1+3\omega}{4}\Big)+ \nonumber \\
    &\frac{3(1-\omega)}{4}\log_2\Big(\frac{1-\omega}{4}\Big)\Big)~.
\end{align}
%for $\omega\!>\!1/3$.
Since $f_\text{de}(1/2)\!=\!0$, Cond.~1~\eqref{eq:c>.5} is met.
Indeed,
$$c^\text{de}:=\sup \{\omega: f_\text{de}(\omega)=0\} \approx 0.747613~.$$ 
For $\omega>c^\text{de}$, we define $F_\text{de} := \ln(f_\text{de})$.
Further,
\begin{align*}
    f'_\text{de}(\omega) &\!=\!
    \frac{3}{4} \log_2 \Big(\frac{1+3 \omega}{1-\omega}\Big)~,\quad f''_\text{de}(\omega) \!=\! \frac{3\log_2 e}{-3w^2 + 2w + 1}\\
    F'_\text{de}(\omega) &= \frac{f'_\text{de}(\omega)}{f_\text{de}(\omega)}~, \quad F''_\text{de}(\omega) = \frac{f''_\text{de}(\omega) f_\text{de}(\omega)\!-\! (f'_\text{de}(\omega))^2}{(f_\text{de}(\omega))^2}~.
\end{align*}

\looseness = -1 In Fig.~\ref{fig:d2F_de}, we see that $F_\text{de}$ has a unique inflection point $c_1^\text{de} \!\approx\! 0.966984$.
In Fig.~\ref{fig:g_de}, we plot ${g_\text{de}(u) \!=\! 2\!-\!u F_\text{de}^{\prime \prime}(u)/(u F_\text{de}^{\prime \prime}(u) \!+\! F_\text{de}^{\prime}(u))\!-\!1/u}$, to show that Cond.~2~\eqref{eq:condFconcave} is met.
Thus, by Thm.~\ref{thm:FuConcave}, the contribution of a route to the objective function in formulation~\eqref{eq:QNUMtransformed} is convex if $f_\text{de}$ is used as its measure of entanglement. \vspace{-9pt}

\begin{figure*}[t!]
\centering
\includegraphics[width=0.95\textwidth]{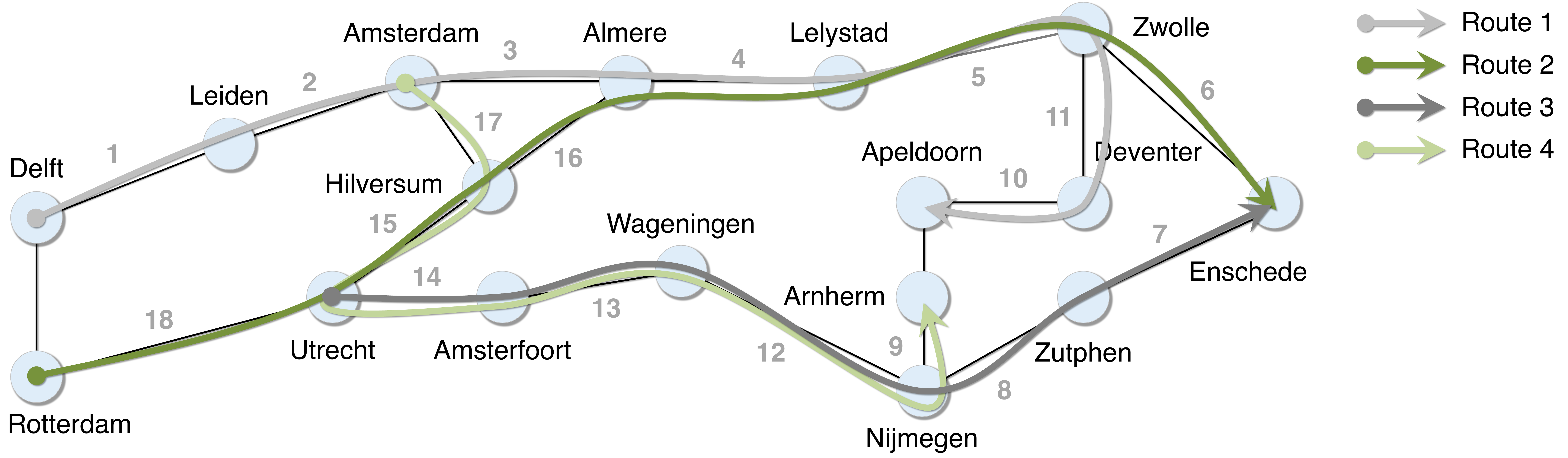}
\caption{\label{fig:surf}%
A subgraph of the SURFnet topology from~\cite{vardoyan2024bipartite}, figure not to scale. Users run QKD on $4$ routes (Tab.~\ref{tab:routes}) using $18$ links annotated here. The length of a link determines its transmissivity (Tab.~\ref{tab:links}). Consistent with current hardware capabilities, entanglement generation is assumed to be attempted every $T_j \!=\!10^{-3}$ seconds and the non-fibre induced inefficiencies coefficient is $\kappa_j \!=\! 0.1$ for each link $j \in [18]$. Optimal allocations are provided in Tab.~\ref{tab:optRateFid}.}%
\vspace{-10pt}
\end{figure*}

\subsection{Negativity}\label{sec:negativity}
In~\cite{vardoyan2023quantum}, it was already shown that negativity allows for convex formulation via~\eqref{eq:QNUMcanonical}.
However, this example is relevant to QNUM problems in networks where one route's entanglement measure is defined by negativity, and another route's entanglement measure requires the variable transformation in~\eqref{eq:cov} to achieve a convex formulation. 

For a Werner state with Werner parameter $\omega$, negativity can be expressed as
\begin{align}\label{eq:neg}\vspace{-8pt}
    f_\text{neg}(\omega) \!=\! \max\Big(0,\frac{3 \omega-1}{4}\Big)~.
\end{align}
In~\cite[App. A]{vardoyan2023quantum}, it was shown that the function 
\begin{align*}\vspace{-8pt}
    J(\Vec{x},\Vec{w}) := -\!\ln x_i \!-\! \ln\Big(f_\text{neg}\Big(\prod\limits_{j=1}^l w_j^{a_{ji}}\Big)\Big)
\end{align*}
is convex on $S' \!:=\! \{(\vec{x},\vec{w})\!:\vec{0} \!\prec\! \vec{x},\vec{0} \!\prec \!\vec{w},\prod_j w_j\!>\!1/3\}$.
This implies that $J(\Vec{x},\Vec{w})$ is also convex on the convex subdomain ${S''\!:=\!\{(\vec{x},\vec{w})\!\in\! S':w_j \!\le\! 1\!-\!\langle A_j,\vec{x} \rangle/d_j~ \forall j\}}$.
Since extended-value extensions of convex functions are convex,
\begin{align*}\vspace{-5pt}
    \Tilde{J}(\vec{x},\vec{w}):=\begin{cases}
     J(\Vec{x},\Vec{w})~,\quad (\Vec{x},\Vec{w})\in S''\\
     \infty,~\quad \text{otherwise}
    \end{cases}
\end{align*}
is convex as well.
Therefore,
\begin{align*}\vspace{-5pt}
    \lbar{J}(\Vec{x}) &= \inf\limits_{\vec{w}\in \mathbb{R}^l} \Tilde{J}(\Vec{x},\Vec{w}) \\
    & = -\!\ln x_i \!-\! \ln\Big(f_\text{neg}\Big(\prod\limits_{j=1}^l \Big( 1-\frac{\langle A_j,\vec{x} \rangle}{d_j}\Big)^{a_{ji}}\Big)\Big)
\end{align*}
is convex by~\cite[Sect. 3.2.5]{boyd2004convex}. 
Note that we have used the fact that $-\ln( f_\text{neg})$ is a decreasing function in the last step.
Thus, $f_\text{neg}$ satisfies the hypothesis of Prop.~\ref{prop:FxToFy} and $\lbar{J}(e^{\vec{y}})$ is convex in $\vec{y}$.
That is, a route with $f_i \!=\! f_\text{neg}$ allows for a convex formulation via~\eqref{eq:QNUMtransformed} even after performing the change of variable~\eqref{eq:cov}. \vspace{-8pt}

\subsection{Success Probability of Teleportation}\label{sec:succTel}
%
% In our teleportation setup, we assume that each qubit at the end nodes is affected by a depolarising noise~\cite{nielsenChuang}:
% \begin{align}\label{eq:noiseModel}
%     &\mathcal{N}_t(\rho) \!= (1\!-\!p(t))\rho
%     \!+\! p(t)~ \nicefrac{\mathbb{I}_2}{2},
% \end{align}
% where the density matrix $\rho$ (resp. $\mathcal{N}_t(\rho)$ ) denotes the state of the qubit at time $0$ (resp. time $t$).
% Further, $p(t) = 1-e^{-t/t_c}$ and $t_c$ is the memory lifetime constant characterising the depolarising effect.
The success probability (or fidelity) of teleportation with a Werner state with Werner parameter $\omega$ as a shared entanglement resource is given by
%
% \begin{align}\label{eq:succ}
%     f_\text{succ}(\omega) \!=\! \frac{2-(1-\omega) e^{-t/T_C}}{2}~,\thickspace \omega>0~,
% \end{align}
%
\begin{align}\label{eq:succ}\vspace{-5pt}
    f_\text{succ}(\omega) \!=\! \frac{1+\omega}{2}~,\quad 0 \le \omega\le 1~.
\end{align}
%where $t$ is the classical communication time between the respective nodes; see Appendix for details.

Observe that $f_\text{succ}(0) \!=\! 1/2$, i.e., it does not satisfy assumption~\ref{A:utility}, which requires $f_i(0)\!=\!0$.
However, the corresponding QNUM formulation~\eqref{eq:QNUMtransformed} is similar and the contribution from the $i$-th route is
\begin{align}\label{eq:QNUMsucc}\vspace{-5pt}
\begin{aligned}
     & -y_i - \ln\Big(f_\text{succ}\Big(\prod\limits_{j=1}^l \Big( 1\!-\!\frac{\langle A_j,e^{\vec{y}} \rangle}{d_j}\Big)^{a_{ji}}\Big) \\
    \text{s.t.}&\quad  0\le\frac{\langle A_j,e^{\vec{y}} \rangle}{d_j}<1~, \thickspace j \in [l]~.
\end{aligned}
\end{align}
%The feasible region of the problem is given by
% $$\bar{T}_j = \{\Vec{y} \!\in\! \mathbb{R}^r\!\!:\langle A_j,e^{\vec{y}} \rangle \!\le\!d_j\}~, \quad j \in [l]~.$$
% which is shown in Thm.~\ref{prop:convexDomain} to be convex.

The objective function in~\eqref{eq:QNUMsucc} is not convex in general but if we restrict the end-to-end Werner parameter to $(1/2,1]$, it becomes convex.
That is, we require
\begin{align}\label{eq:succDomain}
\begin{aligned}
    &0<\frac{\langle A_j,e^{\vec{y}} \rangle}{d_j}<1~, \thickspace j \in [l]~,\\
    \text{and} \quad & \nicefrac{1}{2} <\prod\limits_{j=1}^l \Big( 1-\frac{\langle A_j,e^{\vec{y}} \rangle}{d_j}\Big)^{a_{ji}}~.
\end{aligned}
\end{align}

The region corresponding to~\eqref{eq:succDomain} is convex by Thm.~\ref{prop:convexDomain} and the objective function in~\eqref{eq:QNUMsucc} is convex on this domain due to the following: (i) we can write the objective function as 
 ${-y_i\!-\!\ln((1\!+\! u_i(\vec{y}))/2)}$,
(ii) $u_i(\vec{y})$ is concave on ${\{\vec{y}\!:u_i(\vec{y})\!>\!1/2\}}$ by Prop.~\ref{prop:uiConcave} (see the proof in Sect.~\ref{sec:proofs}) and hence ${(1\!+\! u_i(\vec{y}))/2}$ is concave,
(iii) logarithm is concave and increasing implying that ${\ln(1\!+\!u_i(\vec{y}))/2}$ is concave, 
(iv) $-y_i$ is convex in $\vec{y}$.
% The proof is similar to that of Thm.~\ref{prop:convexDomain}; see Appendix for details.

% Since $u_i(\vec{y})$ is concave on each $T_j$ by Prop.~\ref{prop:uiConcave} and $\gamma_i\!>\!0$, so is $(\delta_i+\gamma_i u_i(\vec{y}))/2$.

%
% \begin{align*}
%     &-\!\ln e^{y_i} \!-\! \ln\Big(f_\text{succ}\Big(\prod\limits_{j=1}^l \Big( 1\!-\!\frac{\langle A_j,e^{\vec{y}} \rangle}{d_j}\Big)^{a_{ji}}\Big) \\
%     = & -y_i-\ln\Big( \frac{\delta_i+\gamma_i u_i(\vec{y})}{2}\Big)~,
% \end{align*}
% %
% with $\delta_i \!=\! 2\!-e^{-t/T_C}$ and $\gamma_i \!=\! e^{-t/T_C}$ for the $i$-th route.
% Since $-y_i$ is a convex function of $\vec{y}$, we only need to show that

\section{Numerical Example}\label{sec:eval}
\everypar{\looseness=-1}
In this section, we work out an example on a subgraph~\cite{vardoyan2024bipartite} derived from the network topology of SURFnet, the national research network of the Netherlands.
We assume that each node in the network is also equipped to serve as a repeater and can support multiple routes.
We then consider the scenario if certain pairs of nodes were to perform QKD between themselves on this real-world fibre network.
The corresponding routes and the relevant link IDs are shown in Fig.~\ref{fig:surf}.
In particular, we have the following four routes:
\begin{table}[h!]
\centering
\begin{tabular}{c l c}
Route ID ($i$) & End nodes & Links \\ \hline \hline
1 & (Delft, Apeldoorn) & $(1,2,3,4,5,11,10)$\\ 
2 & (Rotterdam, Enschede) & $(18,15,16,4,5,6)$\\ 
3 & (Utrecht, Enschede) & $(14,13,12,8,7)$\\ 
4 & (Amsterdam, Arnherm) & $(17,15,14,13,12,9)$ \\ 
\hline
\end{tabular}
\caption{QKD routes and incident links; see Fig.~\ref{fig:surf}.}
\label{tab:routes}\vspace{-8pt}
\end{table}
\begin{table}[h!]
\centering
\begin{tabular}{c l c c}
Link ID ($j$) & Link & Length (km) & $d_j$ \\ \hline \hline
1 & (Delft, Leiden) & 30.6 & 89.84 \\ 
2 & (Leiden, Amsterdam) & 60.4 & 53.79 \\ 
3 & (Amsterdam, Almere) & 38.9 & 77.47 \\ 
4 & (Almere, Lelystad) & 44.2 & 69.44 \\ 
5 & (Lelystad, Zwolle) & 47.7 & 65.12 \\ 
6 & (Zwolle, Enschede) & 78.7 & 40.76 \\ 
7 & (Zutphen, Enschede) & 60 & 54.17 \\ 
8 & (Nijmegen, Zutphen) & 58.1 & 56.25 \\ 
9 & (Nijmegen, Arnhem) & 25.7 & 99.02 \\ 
10 & (Apeldoorn, Deventer) & 24.4 & 100.98 \\ 
11 & (Deventer, Zwolle) & 44.7 & 68.75 \\ 
12 & (Wageningen, Nijmegen) & 66.3 & 49.35 \\ 
13 & (Amersfoort, Wageningen) & 62.5 & 52.40 \\ 
14 & (Utrecht, Amersfoort) & 33.8 & 84.63 \\ 
15 & (Utrecht, Hilversum) & 36.7 & 80.54 \\ 
16 & (Hilversum, Almere) & 35.4 & 82.41 \\ 
17 & (Amsterdam, Hilversum) & 30.2 & 90.52 \\ 
18 & (Rotterdam, Utrecht) & 70 & 46.82 \\ \hline
\end{tabular}
\caption{Relevant links from Fig.~\ref{fig:surf} and derived $d_j$'s.}
%\vspace{-12pt}
\label{tab:links}\vspace{-10pt}
\end{table}

Since all routes perform QKD, we assume that the relevant measure of entanglement for expressing the utility of each route is the secret key fraction, i.e., $f_i \!=\! f_\text{sk},~i \in [4]$; see~\eqref{eq:sk}.
To cast the QNUM problem in this network to formulation~\eqref{eq:QNUMtransformed}, we only need the constants $d_j$'s for each relevant link.

Recall from Assumption~\ref{A:attemptFreq} that $d_j = 3 \kappa_j \eta_j/ 2T_j$, where $\kappa_j, \eta_j$ and $T_j$ denote the constant reflecting inefficiencies other than photon loss in the fibre, the transmissivity and the frequency of the entanglement generation attempt for the $j$-th link, respectively.
Similar to an example in~\cite{vardoyan2023quantum}, we assume that $\kappa_j = 0.1$ and $T_j = 10^{-3}$ seconds for all $18$ links.
The tranmissivities can be calculated as $\eta_j = 10^{-0.02 L_j}$, where $L_j$ is the length of the $j$-th link in km.
The final values of the constants $d_j$'s are provided in Tab.~\ref{tab:links}.

We now solve the QNUM problem via formaulation~\eqref{eq:QNUMtransformed} using MATLAB's \texttt{fmincon}.
The convex nature of the problem ensures that the output $y_i$'s are indeed globally optimal.
The optimal rate allocations and link-level Werner parameters are then computed as $x_i \!=\! e^{y_i}$ and using~\eqref{eq:wFromY}, respectively.
This prescribes the optimal setting for the link-level generation rates and the fidelities.
The resulting end-to-end Werner parameters and fidelities for the optimal allocation are as follows. 
\begin{table}[h!]
\centering
\begin{tabular}{c c c c c}
Route ID & $y_i$ & Rate ($x_i$) in & Werner param.  & Fidelity \\
($i$) & & pairs/sec. & ($u_i$) & \\
\hline \hline
1 & -0.1530 & 0.8581 & 0.8991 & 0.9243 \\
2 & -0.2850 & 0.7520 & 0.8950 & 0.9212 \\
3 & -0.2523 & 0.7770 & 0.8994 & 0.9245 \\
4 & -0.3268 & 0.7213 & 0.8945 & 0.9209 \\
\hline
\end{tabular}
\caption{Optimal rate and fidelity allocations.}
\label{tab:optRateFid} \vspace{-13pt}
\end{table}

%\sk{Teleportation example: The optimisation assigns the lowest possible fidelity to the routes as $f_\text{succ} \!\ge\! 0.5$. }
\section{Proofs of Results}\label{sec:proofs}
\everypar{\looseness=-1}
\begin{proof}[Proof of Thm.~\ref{prop:convexDomain}]
    We make use of the fact that sublevel (resp. superlevel) sets of convex (resp. concave) functions are convex and the feasible set $S$ can be expressed as an intersection of convex sets.
    Recall from~\eqref{eq:feasibleReg} that 
    \begin{align*}
    \begin{aligned}
        T_j \!:= &\{\Vec{y} \!\in\! \mathbb{R}^r\!\!:\langle A_j,e^{\vec{y}} \rangle \!<\!d_j\}, \thickspace j \in [l], ~\thickspace 
    T \!:= \cap_{j \in [l]} T_j~, \\
    S_i \!:= &\{\Vec{y} \!\in\! T\!\!:u_i(\Vec{y})\!>\!c^{(i)}\!\}, \thickspace i \in [r], ~\thickspace
    S \!:= \cap_{i \in [r]} S_i~.
    \end{aligned}
    \end{align*}
    By convexity of the exponential function and the fact that $a_{ji} \ge 0$, $\langle A_j,e^{\vec{y}} \rangle$ is a convex function of $\vec{y}$. 
    Thus, $T_j$'s, being sublevel sets of convex functions, are convex.
    Further, $T$ being the intersection of $T_j$'s is also convex.
    
    Since $c^{(i)}>0$, we now consider the following set:
    \begin{align*}
        \Tilde{S}_i := \{\Vec{y}\!\in\! \mathbb{R}^r:\ln(u_i(\Vec{y}))\!>\!\ln c^{(i)}\}~.
    \end{align*}
    We will show that $\Tilde{S}_i$ is convex. Observe that
    $$\ln(u_i(\Vec{y})) = \sum\limits_{j=1}^l a_{ji} \ln \Big( 1-\frac{\langle A_j,e^{\vec{y}} \rangle}{d_j}\Big)~,$$
    where $a_{ji}\ge 0$.
    Thus, to prove that $\Tilde{S}_i$ is convex, it suffices to show that each $\ln (1-\langle A_j,e^{\vec{y}} \rangle /d_j)$ is concave.
    To this end, we consider $\vec{y}_1,\vec{y}_2 \in \Tilde{S}_i$ and ${\vec{y}_t = t\vec{y}_1+(1-t)\vec{y}_2}$, $t \in (0,1)$.
    By convexity of $\langle A_j,e^{\vec{y}} \rangle$
    \begin{align}
        &\langle A_j,e^{\vec{y}_t} \rangle \le t \langle A_j,e^{\vec{y}_1} \rangle+ (1-t) \langle A_j,e^{\vec{y}_2} \rangle \\
        \overset{d_j>0}{\implies} &1-\frac{\langle A_j,e^{\vec{y}_t} \rangle}{d_j} \ge t \Big( 1-\frac{\langle A_j,e^{\vec{y}_1} \rangle}{d_j}\Big) + \nonumber \\
        & \qquad \qquad (1-t)\Big(1-\frac{\langle A_j,e^{\vec{y}_2} \rangle}{d_j}\Big)\\
        \implies & \ln\Big(1-\frac{\langle A_j,e^{\vec{y}_t} \rangle}{d_j}\Big) \ge \ln \Big( t \Big(1-\frac{\langle A_j,e^{\vec{y}_1} \rangle}{d_j}\Big) + \nonumber \\
        & \qquad \qquad (1-t)\Big( 1-\frac{\langle A_j,e^{\vec{y}_2} \rangle}{d_j}\Big)\Big) \label{eq:step1}\\
        \implies & \ln \Big( 1-\frac{\langle A_j,e^{\vec{y}_t} \rangle}{d_j}\Big) \ge t \ln \Big(1-\frac{\langle A_j,e^{\vec{y}_1} \rangle}{d_j}\Big)+ \nonumber\\
        & \qquad \qquad (1-t)\ln \Big(1-\frac{\langle A_j,e^{\vec{y}_2} \rangle}{d_j}\Big)~,\label{eq:step2}
    \end{align}
    where we have used monotonicity and concavity of logarithm in~\eqref{eq:step1} and~\eqref{eq:step2}, respectively.
    
    Since $S$ can be rewritten as an intersection of convex sets:  $S = (\cap_{i \in [r]} \Tilde{S}_i) \cap T$, it must be convex.
\end{proof}
The following lemma is crucial for showing that $u_i(\vec{y})$ is concave on $S_i$. 
\begin{lemma}\label{lem:forConcavity}
    Let $0< b_\iota < 1,~n \ge 2$ and $0<t<1$.
    Then,
    \begin{align}\label{eq:forConcavity}
        \sup\limits_{\prod_{\iota=1}^n b_\iota \ge t} \big(\sum_{\iota=1}^{n-1} \nicefrac{1}{b_\iota} \big) = n-2+\nicefrac{1}{t}~.
    \end{align}
\end{lemma}
\begin{proof}
    We again use the change of variable technique~\cite{boyd2007tutorial}:
    $e^{p_\iota} \!=\! b_\iota,~p_\iota<0,~\iota \in [n].$
    Then,
    $$\sup\limits_{\prod_{\iota=1}^n b_\iota \ge t,\thinspace 0<b_\iota<1 \thinspace \forall \iota} \!\!\big(\sum_{\iota=1}^{n-1} \nicefrac{1}{b_\iota} \big) \!=\! \sup\limits_{\sum_{\iota=1}^n p_\iota \ge \ln t,\thinspace p_\iota<0 \thinspace \forall \iota} \!\!\big(\sum_{\iota=1}^{n-1} \!e^{-p_\iota} \big).$$
    We prove by induction that 
    \begin{align}\label{eq:covExp}
        \sup\limits_{\sum_{\iota=1}^n p_\iota \ge \ln t,\thinspace p_\iota<0 \thinspace \forall \iota} \big(\sum_{\iota=1}^{n-1} e^{-p_\iota} \big) = n-2+\nicefrac{1}{t}~.
    \end{align}
    To show~\eqref{eq:covExp} for $n=2$, we observe that maximising the function on the LHS reduces to minimising $p_1$ in the feasible set $\{(p_1,p_2): p_1+p_2 \ge \ln t,~ p_1,p_2<0\}$ as the function $e^{-p_1}$ is decreasing in $p_1$.
    Direct substitution gives
    $$\sup\limits_{\sum_{\iota=1}^2 p_\iota \ge \ln t,\thinspace p_\iota<0 \thinspace \forall \iota} e^{-p_1} = \nicefrac{1}{t}~,$$
    proving the assertion for $n=2$.
    Assuming that the hypothesis holds for $n=k-1$,
    \begin{align}
        &\sup\limits_{\sum_{\iota=1}^k p_\iota \ge \ln t,\thinspace p_\iota<0 \thinspace \forall \iota} \big(\sum_{\iota=1}^{k-1} e^{-p_\iota} \big) \\
        = & \sup\limits_{\ln t \le p<0}~\sup\limits_{p_1 = p,\thinspace \sum_{\iota=2}^{k} p_\iota \ge \ln t-p, \thinspace p_\iota<0 \thinspace \forall \iota} \big(e^{-p}+\sum_{\iota=2}^{k-1} e^{-p_\iota} \big) \\
        = & \sup\limits_{\ln t \le p<0} \big(e^{-p}+ ~\sup\limits_{\sum_{\iota=2}^{k} p_\iota \ge \ln t-p, \thinspace p_\iota<0 \thinspace \forall \iota} \sum_{\iota=2}^{k-1} e^{-p_\iota} \big) \\
        = & \sup\limits_{\ln t \le p<0} \big(e^{-p}+k-3+e^{p-\ln t}\big) \\
        = &~ k-3 +\sup\limits_{\ln t \le p<0} \big(e^{-p}+e^{p-\ln t}\big)\\
        = &~ k-2+\nicefrac{1}{t}~\label{eq:lemma1induct},
    \end{align}
    where~\eqref{eq:lemma1induct} follows from considering the convex function $e^{-x}\!+\!e^{x+\lambda}$ over $x \!\in\! [-\lambda,0)$ and observing that its supremum is attained at $-\lambda$.
    Thus, the assertion is true for $n \ge 2$. 
\end{proof} 
To prove Thm.~\ref{thm:FuConcave}, we need a similar result.
\begin{lemma}\label{lem:forConcavityF}
    Let $0 \!<\! b_\iota \!<\! 1,~0\!<\!\beta \!<\! 1,~n \!\ge\! 2$ and $0 \!<\! t \!<\! 1$.
    Then,
    \begin{align}\label{eq:forConcavity2}
        \sup\limits_{\prod_{\iota=1}^n b_\iota = t} \big(\nicefrac{\beta}{b_1}+\sum_{\iota=2}^n \nicefrac{1}{b_\iota} \big) = n-2+\beta+\nicefrac{1}{t}~.
    \end{align}
\end{lemma}
\begin{proof}
    For $n=2$, $\beta\!<\!1$ immediately gives
    \begin{align*}
        \sup\limits_{~b_1 b_2 = t,~ 0<b_\iota<1} \big(\nicefrac{\beta}{b_1}+\nicefrac{1}{b_2} \big) = \beta+\nicefrac{1}{t}~.
    \end{align*}
    The induction step follows similarly to Lemma~\ref{lem:forConcavity} where we let
    $$e^{p_\iota} = b_\iota~, \quad p_\iota<0,~ \iota \in [n]~.$$
    Assuming~\eqref{eq:forConcavity2} holds for $n = k-1$,
    \begin{align}
        &\sup\limits_{\prod_{\iota=1}^k b_\iota = t} \big(\nicefrac{\beta}{b_1}+\sum_{\iota=2}^{k} \nicefrac{1}{b_\iota} \big) \\
        = & \sup\limits_{\sum_{\iota=1}^k p_\iota = \ln t,\thinspace p_\iota<0 \thinspace \forall \iota} \big(\beta e^{-p_1} +\sum_{\iota=2}^{k} e^{-p_\iota} \big) \\
        = & \sup\limits_{\ln t \le p<0}~\sup\limits_{p_k = p,\thinspace \sum_{\iota=1}^{k-1} p_\iota = \ln t-p, \thinspace p_\iota<0 \thinspace \forall \iota} \big(\beta e^{-p_1}+ \nonumber\\
        & \qquad \sum_{\iota=2}^{k-1} e^{-p_\iota} + e^{-p}\big) \\
        = & \sup\limits_{\ln t \le p<0} \big(e^{-p}+\sup\limits_{\sum_{\iota=1}^{k-1} p_\iota = \ln t-p, \thinspace p_\iota<0 \thinspace \forall \iota} \big(\beta e^{-p_1}+ \nonumber\\
        & \qquad \sum_{\iota=2}^{k-1} e^{-p_\iota} \big)\big) \\
        = &~k-3 +\beta +\sup\limits_{\ln t \le p<0} \big(e^{-p}+e^{p-\ln t}\big) \\
        = &~k-2+\beta+ \nicefrac{1}{t}~,
    \end{align}
    where the last step involves computing the maxima of the function $e^{-x}+e^{x+\lambda}$ over $x \in [-\lambda,0)$ as in Lemma~\ref{lem:forConcavity}.
    Thus,~\eqref{eq:forConcavity2} holds for $n \ge 2$.
\end{proof}
\begin{proof}[Proof of Prop.~\ref{prop:uiConcave}]
    Our proof is based on two facts: (i) since $u_i(\vec{y})$ is twice-differentiable, its Hessian $H^{(i)}$ is symmetric by~\cite[(8.12.3)]{dieudonne1970foundations} and has real eigenvalues, (ii) each diagonal element of $H^{(i)}$ dominates the absolute sum of the non-diagonal entries of the corresponding row on $S_i$.
    We can thus apply Gershgorin's circle theorem~\cite{zbMATH02560682} 
    which says that the eigenvalues of a matrix are contained in the circles with centres as diagonal elements and the respective absolute sum of off-diagonal elements as radii.
    In this case, this would imply that the eigenvalues of $H^{(i)}$ are non-positive on $S_i$, which is equivalent to concavity of $u_i(\vec{y})$.
    
    To establish fact~(ii), we observe that the sign of the row-sums of $H^{(i)}$ is the opposite of the sign of sums of the form ${n-\sum_{j \ne j',~ j \in [n]} 1/w_{j}(\vec{y})}$ for some $n \in \mathbb{N}$ and $j' \in [n]$.
    If ${0\!<\!w_{j}(\vec{y})\!<\!1}$ and ${\prod_{j \in [n]} w_{j}(\vec{y})\!>\!1/2}$, such sums are non-negative as $\sum_{j \ne j',~ j \in [n]} 1/w_{j}(\vec{y})$ cannot exceed $n$ (Lemma~\ref{lem:forConcavity}).
    The details are as follows.

    We suppose that the $i$-th route passes through $n$ links where $n\le l$.
    Without loss of generality, we can number these links as $1,2,\dots,n$.
    Then for $k \in [r]$,
    \begin{align}\label{eq:delUdelY}
        \frac{\partial u_i(\vec{y})}{\partial y_k} 
        \!=\! \frac{\partial }{\partial y_k} (\prod\limits_{j=1}^n w_j(\vec{y})) 
        \!=\! \sum\limits_{j=1}^n \underbrace{\frac{\partial w_j(\vec{y})}{\partial y_k} \!\prod\limits_{j' \in [n]\setminus\{j\}} w_{j'}(\vec{y})}_{v^{(i)}_{jk}(\vec{y})}.
    \end{align}
    From now on, we frequently drop the argument $\vec{y}$ when it is clear from context. 
    Note that $v^{(i)}_{j k} = 0$ iff $a_{jk} = 0$, i.e., the $k$-th route does not pass through the $j$-th link.
    Specifically,
    \begin{align}\label{eq:defVjk}
    \begin{aligned}
        v^{(i)}_{jk} = -\frac{a_{jk} e^{y_k}}{d_j}\prod\limits_{j' \in [n]\setminus\{j\}} w_{j'}~.
    \end{aligned}
    \end{align}
    Also,
    \begin{align}
        &\sum_{m=1}^r \frac{\partial w_{j'}}{\partial y_m} = \sum_{m=1}^r \frac{-a_{j'm} e^{y_m}}{d_{j'}} = w_{j'}-1  \label{eq:sumPartials}\\
        \text{and} \quad &\frac{\partial}{\partial y_m} \big(-\frac{e^{y_k}}{d_j}\big) = -\frac{e^{y_k}}{d_j} \mathbbm{1}_{m=k}~.
    \end{align}
    This leads to the following expression for the Hessian:
    \begin{align*}
        & H^{(i)}_{kk}(\vec{y}) = \sum_{j=1}^n \frac{\partial v^{(i)}_{jk}}{\partial y_k}  = -\sum_{j=1}^n \frac{a_{jk}e^{y_k}}{d_j} \prod\limits_{j' \in [n]\setminus\{j\}} w_{j'} \Big( 1+ \\
        & \qquad \sum_{j'' \in [n]\setminus\{j\}} \frac{\partial w_{j''}}{\partial y_k} \frac{1}{w_{j''}}\Big)~,\\
        &H^{(i)}_{km}(\vec{y}) = H^{(i)}_{mk}(\vec{y}) = -\sum_{j=1}^n \frac{a_{jk}e^{y_k}}{d_j} \prod\limits_{j' \in [n]\setminus\{j\}} w_{j'} \Big( \\
        & \qquad \sum_{j'' \in [n]\setminus\{j\}} \frac{\partial w_{j''}}{\partial y_m} \frac{1}{w_{j''}}\Big)~, \quad m \in [r]\setminus{k}~.
    \end{align*}
    % Since $w_{j'}$ is a continuous function of $\vec{y}$ and $w_{j'}>0$, so is $1/w_{j'}$, and hence, $H^{(i)}_{mk}(\vec{y})$.
    % By symmetry of partial derivatives~\cite[Thm. 9.41]{rudin1964principles}, $H^{(i)}_{km}(\vec{y}) = H^{(i)}_{mk}(\vec{y})$.
    % Thus, all eigenvalues of $H^{(i)}$ are real.
    Since $\nicefrac{\partial w_j}{\partial y_s}\le 0~ \forall j,s$, $H^{(i)}_{km}(\vec{y}) \!\ge\! 0$ for $k \ne m$.
    Therefore,
    \begin{align}\label{eq:Hess}
    \begin{aligned}
         &H^{(i)}_{kk}(\vec{y})+\sum_{m \in [r] \setminus k} \big|H^{(i)}_{km}(\vec{y}) \big| \\
        =& \thinspace H^{(i)}_{kk}(\vec{y})+\sum_{m \in [r] \setminus k} H^{(i)}_{km}(\vec{y}) \\
        \overset{\eqref{eq:defVjk}}{=} &\sum_{j=1}^n v^{(i)}_{jk} \Big( 1+ \!\!\sum_{j'' \in [n]\setminus\{j\}}~ \frac{1}{w_{j''}}\Big( \frac{\partial w_{j''}}{\partial y_k} + \sum_{m \in [r] \setminus k} \frac{\partial w_{j''}}{\partial y_m} \Big)\Big)\\
        = & \sum_{j=1}^n v^{(i)}_{jk} \Big( 1+ \!\!\sum_{j'' \in [n]\setminus\{j\}}~ \frac{1}{w_{j''}} \sum_{m=1}^r \frac{\partial w_{j''}}{\partial y_m} \Big)\\
        \overset{\eqref{eq:sumPartials}}{=} & \sum_{j=1}^n v^{(i)}_{jk} \Big( 1+ \!\!\sum_{j'' \in [n]\setminus\{j\}}~ \frac{w_{j''}-1}{w_{j''}} \Big) \\
        = & \sum_{j=1}^n v^{(i)}_{jk} \Big(n-\sum_{j'' \in [n]\setminus\{j\}}~ \nicefrac{1}{w_{j''}} \Big)~.
    \end{aligned}
    \end{align}
    Recall that on $S_i$, ${0\!<\!w_{j''}\!<\! 1}$ for all $j'' \in [n]$ and ${\{\prod_{j'' \in [n]}w_{j''}\!>\! c^{(i)}\}}$.
    We can thus apply Lemma~\ref{lem:forConcavity} to see that
    \begin{align*}
        n-\!\!\sum_{j'' \in [n]\setminus\{j\}} \!\!\nicefrac{1}{w_{j''}} &\ge n-\!\!\sup\limits_{\prod\limits_{j'' \in [n]}w_{j''}\ge c^{(i)}} \Big(\sum_{j'' \in [n]\setminus\{j\}} \!\nicefrac{1}{w_{j''}} \Big) \\
        &= 2-\nicefrac{1}{c^{(i)}}~,    
    \end{align*}
    which is non-negative for $c^{(i)} \ge 1/2$.
    Also, observe from~\eqref{eq:defVjk} that $v^{(i)}_{jk} \le 0$, implying that the RHS of~\eqref{eq:Hess} is non-positive.
    Therefore, all eigenvalues of $H^{(i)}$ are non-positive on $S_i$ due to Gershgorin's circle theorem~\cite{zbMATH02560682}.
\end{proof}

\begin{proof}[Proof of Thm.\ref{thm:FuConcave}]
    Let us denote by $D^2F$ the Hessian of ${F_i(u_i(\vec{y}))= \ln f_i(u_i(\vec{y}))}$.
    We will again exploit the symmetry of the Hessian and use Gershgorin's circle theorem~\cite{zbMATH02560682} to show that the eigenvalues of $D^2F$ are non-positive on $S_i$.
     
    To prove symmetry, we observe that since $u_i(\vec{y})$ is twice-differentiable and $f_i$ is twice-differentiable by assumption, so is $f_i(u_i(\vec{y}))$.
    Further, $f_i(u_i(\vec{y}))\!>\!0$ on $S_i$.
    Thus, $F_i(u_i(\vec{y}))$ is twice-differentiable on $S_i$ and $D^2F$ is symmetric by~\cite[(8.12.3)]{dieudonne1970foundations}. 
    
    Also, $D^2F$ can be explicitly written as
    \begin{align}\label{eq:hessF}
        D^2F(\Vec{y}) \!=\! F_i''(u_i(\vec{y}))(u'_i(\vec{y}))^T u'_i(\vec{y}) \!+\! F_i'(u_i(\vec{y})) H^{(i)}(\vec{y}),
    \end{align}
    and we consider it separately on the following subsets of $S_i$: ${\{\vec{y}\!\in\!T\!:\! c^{(i)}\!<\!u_i(\vec{y})\!\le\! c_1^{(i)}\}}$,~ ${\{\vec{y}\!\in\!T\!:\! c_1^{(i)}\!<\!u_i(\vec{y})\!<\!1\}}$.
    This is because $F_i''(u) \!\le\! 0$ for $u \!\in\! (c^{(i)},c_1^{(i)}]$ and $F_i''(u)\!>\! 0$ for $u \!\in\! (c_1^{(i)},1)$ by assumption.
    Since $(u'_i)^T u'_i$ is positive semidefinite, $F'_i \!\ge\! 0$ (as $f_i$ is increasing and $F_i \!=\! \ln(f_i)$), $H^{(i)}$ is negative semidefinite on $S_i$ (due to the assumption that Cond. 1~\eqref{eq:c>.5} holds) and $F_i'' \!\le\! 0$ on the first subset, $D^2F$ is negative semidefinite on this subset as well. 
    
    We now consider $D^2F$ on ${\{\vec{y}\!\in\!T\!:\! c_1^{(i)}\!<\!u_i(\vec{y})\!<\!1\}}$ (the second subset of $S_i$) where $F_i''(u_i(\vec{y})\!>\!0$.
    Expanding~\eqref{eq:hessF} 
    \begin{align*}
        \begin{aligned}
        D^2 F_{k k}(\vec{y}) & \!=\!F_i^{\prime \prime}\left(u_i(\vec{y})\right)\!\Big(\frac{\partial}{\partial y_k} u_i(\vec{y})\!\Big)^2\!\!+\!F_i^{\prime}\left(u_i(\vec{y})\right) H_{k k}^{(i)}(\vec{y}),\\
        D^2 F_{km}(\vec{y}) & \!=\! D^2 F_{m k}(\vec{y})\\
        & \!=\!F_i^{\prime \prime}(u_i(\vec{y}) \frac{\partial u_i(\vec{y})}{\partial y_m}  \frac{\partial u_i(\vec{y})}{\partial y_k} \!+\! F_i^{\prime}\left(u_i(\vec{y})\right) H_{m k}^{(i)}(\vec{y}),
        \end{aligned}
    \end{align*}
    where $m \in [r]\setminus{k}$.
    %Since $f_i(u_i(\vec{y}))$ is twice continuously differentiable and $f_i(u_i(\vec{y}))>0$ on $\{\vec{y}: u_i(\vec{y})> c_1\}$, $F_i''(u_i(\vec{y}))$ is continuous as well.
    %Recall from the proof of Prop.~\ref{prop:uiConcave} that $H^{(i)}_{km}(\vec{y}) = H^{(i)}_{mk}(\vec{y})$.
    %Thus, by symmetry of partial derivatives~\cite[Thm. 9.41]{rudin1964principles}, $D^2F_{km}(\vec{y}) = D^2F_{mk}(\vec{y})$ on $\{\vec{y}: u_i(\vec{y})> c_1^{(i)}\}$ and all eigenvalues of .
    Since $\nicefrac{\partial u_i}{\partial y_s}\le 0~ \forall s$ (see~\eqref{eq:delUdelY},~\eqref{eq:defVjk}), $H^{(i)}_{mk}(\vec{y}) \ge 0$, $F'_i\ge 0$ and $F''_i \!>\! 0$ on the second subset, $D^2 F_{k m}(\vec{y}) \ge 0$ for $m \in [r]\setminus{k}$.
    Further,
    \begin{align}\label{eq:sumDelUdelY}
    \begin{aligned}
        \sum_{m=1}^r \frac{\partial u_i(\vec{y})}{\partial y_m}
        =&\sum_{j=1}^n \sum_{m=1}^r \frac{-a_{j m} e^{y_m}}{d_j}\!\!\!\! \prod_{j' \in [n]\setminus\{j\}} \!\!\!\!w_{j^{\prime}} \\
        =&\sum_{j=1}^n\left(w_j\!-\!1\right) \!\!\!\!\prod_{j' \in [n]\setminus\{j\}} \!\!\!\!w_{j \prime}\\
        =& u_i \big(n\!-\!\sum_{j=1}^n \nicefrac{1}{w_j}\big).
    \end{aligned}
    \end{align}
    Therefore,
    \begin{align}
        & \quad D^2 F_{k k}(\vec{y})+\sum_{m \in [r] \setminus k} \big|D^2 F_{k m}(\vec{y}) \big| \\
        & = D^2 F_{k k}(\vec{y})+\sum_{m \in [r] \setminus k} D^2 F_{k m}(\vec{y})\\
        & = F^{\prime \prime}_i(u_i(\vec{y})) \frac{\partial u_i(\vec{y})}{\partial y_k} \sum_{m=1}^r \frac{\partial u_i(\vec{y})}{\partial y_m}+F^{\prime}_i(u_i(\vec{y})) \sum_{m=1}^r H_{k m}^{(i)} \\
        & \overset{\eqref{eq:sumDelUdelY}}{=} F^{\prime \prime}_i(u_i(\vec{y})) \frac{\partial u_i(\vec{y})}{\partial y_k} u_i(\vec{y}) \Big(n\!-\!\sum_{j^{\prime}=1}^n \nicefrac{1}{w_{j \prime}}\Big)+ \nonumber\\
        & \qquad F^{\prime}_i(u_i(\vec{y})) \sum_{m=1}^r H_{k m}^{(i)} \\
        & \overset{\eqref{eq:delUdelY},\eqref{eq:Hess}}{=} F^{\prime \prime}_i(u_i(\vec{y})) \Big(\sum_{j=1}^n v_{j k}^{(i)}\Big) u_i(\vec{y}) \Big(n\!-\!\sum_{j^{\prime}=1}^n \nicefrac{1}{w_{j \prime}}\!\Big)+ \nonumber\\
        & \qquad F_i^{\prime}\left(u_i(\vec{y})\right) \sum_{j=1}^n v_{j k}^{(i)}\Big(n-\!\!\!\!\sum_{j'' \in [n]\setminus\{j\}} \!\!\nicefrac{1}{w_{j''}} \!\Big)\\
        & = \sum_{j=1}^n v_{j k}^{(i)}\bigg(\big(u_i F_i^{\prime \prime}(u_i) + F_i^{\prime}(u_i)\big)\Big(n-\!\!\!\sum_{j' \in [n]\setminus\{j\}} \nicefrac{1}{w_{j'}} \Big)- \nonumber\\
        & \qquad u_i F_i^{\prime \prime}(u_i) \nicefrac{1}{w_j}\bigg)\\
        & = \underbrace{\big(u_i F_i^{\prime \prime}(u_i) + F_i^{\prime}(u_i)\big)}_{>0}\sum_{j=1}^n \underbrace{v_{j k}^{(i)}}_{\le 0}\Big(n-\!\!\!\sum_{j' \in [n]\setminus\{j\}} \nicefrac{1}{w_{j'}} ~-\nonumber\\
        & \qquad \underbrace{\frac{u_i F_i^{\prime \prime}(u_i)}{u_i F_i^{\prime \prime}(u_i) + F_i^{\prime}(u_i)}}_{=\beta(u_i)} \frac{1}{w_j}\Big)~.\label{eq:cond2equivalence}
    \end{align}
    \normalfont{
    Recall that $\prod_{j'=1}^n w_{j'} \!=\! u_i$. 
    Also, on the present subset ${c_1^{(i)}\!<\!u_i(\vec{y})\!<\!1}$ and $F_i''(u_i(\vec{y}))\!>\!0$ by definition, meaning that we have $0\!<\!\beta(u_i)\!<\!1$. 
    Thus by Lemma~\ref{lem:forConcavityF},
    \begin{align}
        & n-\!\!\!\sum_{j' \in [n]\setminus\{j\}} \nicefrac{1}{w_{j'}} - \nicefrac{\beta(u_i)}{w_j} \\
        \ge~ & n-\! \sup\limits_{\prod_{j'=1}^n w_{j'} = u_i}\! \Big(\nicefrac{\beta(u_i)}{w_j}+\!\!\!\sum_{j' \in [n]\setminus\{j\}} \nicefrac{1}{w_{j'}} \Big) \\
        \overset{\eqref{eq:forConcavity2}}{=}~ & 2\!-\!\beta(u_i)\!-\!\nicefrac{1}{u_i}~. \label{eq:forCond2}
    \end{align}
    Cond. 2~\eqref{eq:condFconcave} simply says that~\eqref{eq:forCond2} is non-negative on ${\{\vec{y} \!\in\! T\!: u_i(\vec{y})\!>\! c_1^{(i)}\}}$, which via~\eqref{eq:cond2equivalence} implies that 
    $$D^2 F_{k k}(\vec{y})+\sum_{m \in [r] \setminus k} |D^2 F_{k m}(\vec{y})| \le 0~.$$ 
    \looseness = -1 Applying Gershgorin's circle theorem~\cite{zbMATH02560682}, we see that all eigenvalues of $D^2F$ are non-positive if Cond. 2~\eqref{eq:condFconcave} is met.
    
    We have thus shown that Cond. 2~\eqref{eq:condFconcave} is sufficient for $D^2F$ to be negative semidefinite on entire $S_i$, or equivalently, for $F_i(u_i(\vec{y}))$ to be concave.}
\end{proof}
\section{Conclusion}
\everypar{\looseness=-1} In this work, we have focussed on finding conditions under which the QNUM problem can be formulated as a convex problem. 
Like the classical NUM problem, we have shown that the QNUM problem can also be formulated as an optimisation problem solely in terms of rate allocations. 
We then provided a reformulation and sufficient conditions in terms of the relevant entanglement measures for this reformulation to be convex. 
These conditions were shown to hold for previously considered entanglement measures that did not directly admit a convex QNUM formulation. 
The reformulation was shown to preserve convexity, i.e., while attempting to convexify the contribution of a route to the objective function, the reformulation does not render already convex contributions from other routes non-convex.
We also worked out an example where we derived the optimal rate-fidelity allocations if we were to run QKD on a real-world fibre network.
Our findings allow for efficient computation of globally optimal rate and fidelity allocations in an entanglement distribution network supporting diverse applications.

\section*{Acknowledgement}
This work was supported in part by NWO VICI grant VI.C.222.029.
SK thanks Scarlett Gauthier and Kaushik Senthoor for careful reading of an earlier version of the manuscript.

\EOD

\end{document}